# Spontaneous Formation of Stable Capillary Bridges for Firming Compact Colloidal Microstructures in Phase Separating Liquids: A Computational Study

Tian-Le Cheng[*] and Yu U. Wang[†]

Department of Materials Science and Engineering, Michigan Technological University, Houghton, MI 49931, USA

**ABSTRACT:** Computer modeling and simulations are performed to investigate capillary bridges spontaneously formed between closely packed colloidal particles in phase separating liquids. The simulations reveal a self-stabilization mechanism that operates through diffusive equilibrium of two-phase liquid morphologies. Such mechanism renders desired microstructural stability and uniformity to the capillary bridges that are spontaneously formed during liquid solution phase separation. This self-stabilization behavior is in contrast to conventional coarsening processes during phase separation. The volume fraction limit of the separated liquid phases as well as the adhesion strength and thermodynamic stability of the capillary bridges are discussed. Capillary bridge formations in various compact colloid assemblies are considered. The study sheds light on a promising route to *in-situ* (in-liquid) firming of fragile colloidal crystals and other compact colloidal microstructures via capillary bridges.

## INTRODUCTION

A small amount of liquid bridging the wettable or partially wettable surfaces of two solid particles may produce strong adhesion between the particles due to surface tension as well as Laplace pressure.[1-3] Such liquid bridges of pendular shape, so-called capillary bridges, play important roles in various places, such as plastic-like wet sand/soil[4,5] and adhesive toe pads of some small animals.[6] At small length scales capillary forces usually dominate over particle gravity and other surface forces including van der Waals forces.[1, 7-9] Based on capillarity effect, an electronically controlled switchable adhesion device was recently engineered.[10] Most studies focused on capillary bridges formed by liquid in vapor through condensation, while capillary bridges can also be formed by two immiscible liquid phases. If the two liquid phases are insoluble in each other, mechanical agitation needs to be employed to distribute the preferential liquid phase into the tiny gaps between particles to form liquid bridges. Such a method has been used to demonstrate that mixing a small amount of a second insoluble liquid into a colloid suspension can dramatically change its rheology from fluid-like to gel-like.[7] However, mechanical agitation inevitably changes particle arrangements thus cannot be used to introduce capillary bridges into well-formed particle microstructures. An appealing way to introduce capillary bridges is through binary liquid phase separation controlled by composition and temperature.[11] Such a method is expected to be able to stabilize engineered particle microstructures. Nevertheless, a serious concern is the stability of such formed capillary bridges due to liquid clustering phenomena frequently observed in wet particulate materials,[11-13] since coarsening is a general phenomenon in phase separation. Existing relevant theoretical work and molecular level simulations mostly focus on the kinetic process of an isolated capillary bridge in given ambient vapor pressure,[14-17] or the macroscopic bulk phase transition shift due to capillary effect.[18, 19] In the well-studied capillary condensation problem, the equilibrium state of a liquid meniscus is described by the Kelvin equation valid down to a few nanometers of meniscus radius,[20, 21] which links individual radii of curvature of liquid surface to the ambient vapor pressure. Because the ambient vapor pressure is usually constant, the interaction among different menisci is rarely considered. In a two-liquid-phase colloid system, however, a meniscus is strongly interacting with adjacent menisci through inter-liquid diffusion. Such interactions among adjacent capillary bridges and the effects on their stabilities during kinetic growth processes are still inadequately studied.

---

[*] E-mail address: tianlec@mtu.edu
[†] E-mail address: wangyu@mtu.edu

In this work, we focus on multiple mutually interacting capillary bridges spontaneously formed between closely packed colloidal particles in phase separating liquids. Computer simulations are performed based on a recently developed diffuse interface field approach (DIFA).[22] A self-stabilization mechanism is revealed that operates through diffusive equilibrium of two-phase liquid morphologies, where adjacent capillary bridges interact with each other through diffusion in analogy to the liquid interaction in communicating vessels through flow. Such a self-stabilization mechanism automatically stabilizes the spontaneously formed capillary bridges between neighboring colloidal particles during liquid solution phase separation, in contrast to conventional coarsening processes.[23] It renders the capillary bridges desired microstructural stability and uniformity, which are important for a practical route to *in-situ* (in-liquid) firming of fragile colloidal crystals and other compact colloidal microstructures by using capillary bridges.

**COMPUTER SIMULATION METHOD**

1. Diffuse interface field approach (DIFA)

Computer simulations of liquid solution phase separation in the gaps between compact colloidal particles are performed by DIFA. In such a model, the system free energy of a binary liquid solution in the presence of solid particles is described by[22]

$$F = \int \left[ f\left(\{c_\alpha\},\{\eta_\beta\}\right) + \sum_\alpha \frac{1}{2} \kappa_\alpha |\nabla c_\alpha|^2 \right] dV, \qquad (1)$$

where $f(\{c_\alpha\},\{\eta_\beta\})$ is the nonequilibrium local bulk chemical free energy density that defines the thermodynamic properties of a multi-phase system including liquid phases and solid particles. Each fluid phase is represented by a concentration field variable $c_\alpha$, and each solid particle by $\eta_\beta$. For a binary liquid solution (with a miscibility gap) in the presence of multiple solid particles, $f(\{c_\alpha\},\{\eta_\beta\})$ is[22]

$$f(\{c_\alpha\},\{\eta_\beta\}) = A\left[\sum_{\alpha=1}^{2}(3c_\alpha^4 - 4c_\alpha^3) + 6\left(\chi c_1^2 c_2^2 + \sum_\beta \sum_{\alpha=1}^{2}\lambda_\alpha c_\alpha^2 \eta_\beta^2\right)\right]. \quad (2)$$

This Landau-type free energy function is phenomenological in nature to reproduce the energy landscape with energy minima at $\{c_{\alpha=i}=1, c_{\alpha'\neq i}=0, \eta_\beta=0\}$ or $\{c_\alpha=0, \eta_{\beta=j}=1, \eta_{\beta'\neq j}=0\}$, i.e., a spatial position is occupied either by liquid phase $i$ or solid particle $j$. In Eq. (2), the concentration $c_\alpha$ is defined as molar fraction of two liquid phase mixture ($c_1 + c_2 = 1$) for convenience, and $\eta_\beta = 1$ inside the particle $\beta$ and $\eta_\beta = 0$ outside.[24] The gradient terms in Eq. (1) describe the energy contributions from liquid-liquid and liquid-solid interfaces. As a result, all field variables $\{c_\alpha\}$ smoothly transit from 1 to 0, forming diffuse interfaces at both liquid-liquid interfaces and liquid-solid interfaces.[25] Also, $\{\eta_\beta\}$ possess smooth transition interfaces.[24] Inside the liquid ($\eta_\beta = 0$), the free energy function in Eq. (2) describes a double-well potential for binary solution with a miscibility gap. The constant $A$ is an energy scaling coefficient, and the parameters $\chi$, $\lambda_\alpha$ and $\kappa_\alpha$ are used to control the fluid-fluid and fluid-solid interfacial energy densities. Thus, the model is able to simulate colloidal particles of different wettabilities.[22] In particular, $\lambda_1 < \lambda_2$ implies that particles have greater affinity with liquid phase 1, resulting in its preferential surface adsorption. Such a preferential surface adsorption effect assists heterogeneous nucleation of phase separation processes.

The kinetic process of isothermal phase separation is described by Cahn-Hilliard equation

$$\frac{\partial c_\alpha}{\partial t} = \nabla \cdot \left(M_\alpha \nabla \frac{\delta F}{\delta c_\alpha}\right), \quad (3)$$

where $M_\alpha$ is the chemical mobility of liquid species for diffusion.

2. Computational details

A semi-implicit spectral method[26, 27] is employed to solve the Cahn-Hilliard equation, which offers significantly improved efficiency and accuracy comparing to the explicit Euler scheme. The time increment is set to $\Delta t = 0.5l^2/(\Delta HM)$, where $l$ is computational grid size, $\Delta H$ is enthalpy of mixing of the binary solution, and $M_\alpha = M$. In the simulations, solid particles in the compact colloidal microstructures are immobile. Thus, the $\{\eta_\beta\}$ field variables are fixed in space and do not evolve. It is worth noting that a unique capability of DIFA model is to treat motions of arbitrary-shaped colloid particles under influences of various forces, including capillary, steric and electrostatic forces, as previously demonstrated;[22,24] however, the focus of this work is on the spontaneous formation of capillary bridges in compact colloidal microstructures where solid particles are pre-assembled and immobile.

It is worth noting that accumulated round-off errors during numerical solution of diffusion equation over long time (large number of iterations) may generate non-negligible composition drift. In order to avoid error accumulation and conserve mass, a very minor correction is made in the simulations after each time increment: $c_i \leftarrow c_i C_i^0/C_i$, where $C_i^0 = \int c_i(\mathbf{r}, t=0)\, dV$ is the total initial mass of each liquid phase, and $C_i = \int c_i(\mathbf{r}, t)\, dV$. While such a correction is negligibly small for each time increment, it is necessary for simulations over a large number of iteration steps, as in our case where capillary bridge formation during phase separation is very sensitive to the volume fraction of the minor phase.

In the simulations, the parameters $A$ and $\chi$ are set to 1/12 and 1, respectively, and $\lambda_A = 0.5$, $\lambda_B = 2.4$, $\kappa_A = 0.4$, $\kappa_B = 1.0$, which give approximately $3l$ thickness of all diffuse interfaces between 10% and 90% of full equilibrium value, and interface energy (normalized by $\Delta H l$) $\gamma_{AB} = 0.262$ for liquid phase A-liquid phase B, $\gamma_{AS} = 0.108$ for liquid phase A-solid particle, and $\gamma_{BS} = 0.240$ for liquid phase B-solid particle. According to Young's equation, the contact angle of liquid interface at solid surface is $\theta = \cos^{-1}[(\gamma_{BS} - \gamma_{AS})/\gamma_{AB}] \approx 60°$. This contact angle is used in this work. Periodic boundary condition is employed for all simulations. While the DIFA model is formulated for three dimensions, computer simulations reported in this work are performed in two dimensions due to the large amount of computing time required to solve diffusion equation over entire phase separation processes.

**SIMULATION RESULTS AND DISCUSSIONS**

1. Simulation of capillary bridge formation in colloidal crystal

We first study the formation of capillary bridges in an ordered colloidal crystal. The simulation starts from a homogeneous binary liquid solution that undergoes a fast cooling into the miscibility gap, with thermally induced small fluctuations to initiate the phase separation process. It has been known that when the supersaturation reaches a threshold,[28, 29] surface adsorption of the preferential liquid species is strong enough,[29, 30] or the inter-particle distance $D$ is sufficiently small (comparable to the thickness of the wetting film),[16, 18] nucleation of the capillary bridges is virtually barrier-free. In our simulation, the preferential adsorption at particle surfaces leads to heterogeneous nucleation in the narrow gaps between solid particles throughout the system, forming small capillary bridges. Subsequent decomposition of the supersaturated solution leads to growth of these capillary bridges. Most importantly, as shown in Figure 1, the simulation reveals a local stabilization mechanism of the capillary bridges that operates through diffusive equilibrium of two-phase liquid morphologies. Such mechanism renders desired microstructural stability and uniformity to the capillary bridges that are spontaneously formed in colloidal crystals during liquid solution phase separation.

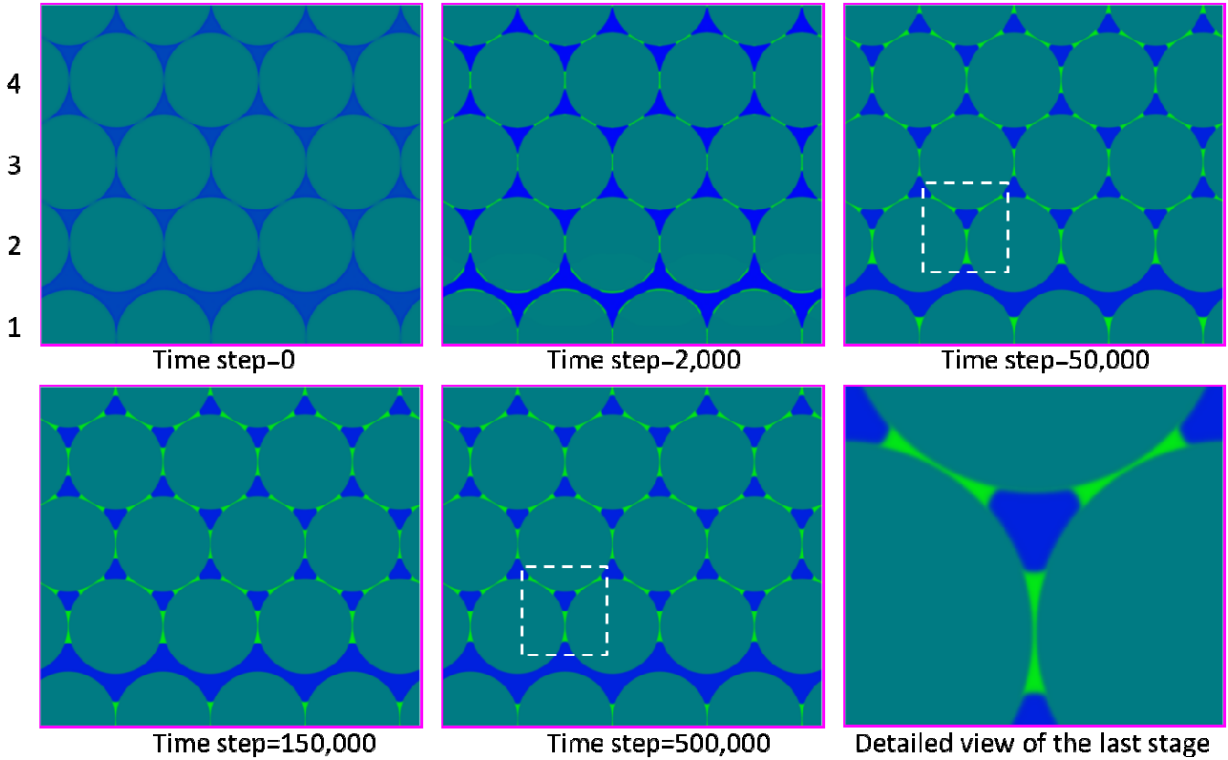

**Figure 1.** Simulated nucleation and growth of capillary bridges formed in colloidal crystal during liquid phase separation process. Colloidal particle volume fraction $f_p = 0.86$. Total fraction of phase A in liquid (fully separated) is $\Phi_A = 1 - \Phi_B = 0.18$. Two-phase liquid morphology reaches equilibrium after 500,000 time steps. Last panel shows the detailed stable capillary bridges at 500,000 time steps. Colors stand for different liquid phases (green, blue) and particles.

Figure 1 shows the simulated time evolution of liquid phase separation process and capillary bridge formation in a colloidal crystal. Space is discretized into 912×816 computational grids, with particle diameter of 226 $l$ to adquately resolve the gap spaces between particles. All particles are in almost contact (numerically there is a small distance of about the diffuse interface thickness betweeb adjacent solid surfaces). An ordered colloidal crystal microstructure is assumed, except for rows 1 and 2 with a broader spacing to intentionally introduce certain degree of structural nonuniformity. It is observed that capillary bridges of liquid phase A (with greater wettability to particle surface) nucleate in the narrowest slits between particles as assisted by the surface adsorption, and the menisci grow gradually. Due to the broader gaps between particles in rows 1 and 2, while surface adsorption occurs on particles (time step 2,000), capillary bridges are not formed there; instead, the abundant liquid component A in this region diffuses away to feed the growth of adjacent capillary bridges that become bigger than furtheraway ones during growth (time step 50,000). In particular, the capillary bridges in row 2 are initially not only bigger but also unsymmetric as compared with those in rows 3 and 4, as highlighted at time step 50,000. However, with phase separation proceeded, these bigger and unsymmetric capillary bridges not only become symmetric but also shrink to the same size of other ones (time step 150,000). In contrast to conventional coarsening processes during phase separation, the capillary bridges do not coarsen or coalesce; instead, they mutually interact through diffusion to balance each other to achieve self-stabilization. When equilibrium is reached in the simulation after time step 500,000, all capillary bridges grow into a uniform size, despite the imperfection (broader spacing between rows 1 and 2) present in the colloidal crystal. The final stable morphology of liquid phase A and solid particles resembles a mortar-brick structure in architecture. In the following sections, we analyze the self-stabilization mechanism of capillary bridges and calculate the adhesion strength of capillary bridge-reinforced colloidal crystal.

2. Analysis of self-stabilization mechanism of capillary bridges

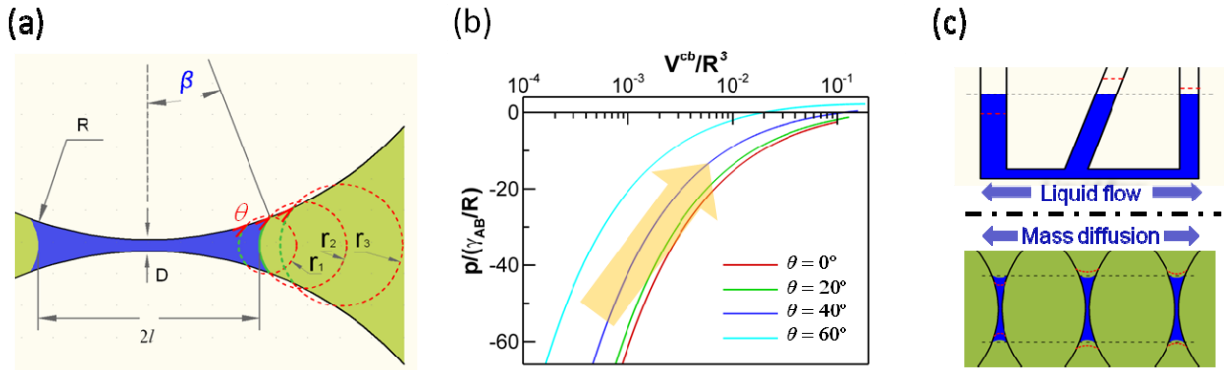

**Figure 2.** (a) Schematic of a capillary bridge between two solid spheres. Blue and green regions are coexisting liquid phases A and B, respectively, with contact angle $\theta < 90°$ (measured in phase A). As small meniscus between spheres grows outward, its radius of curvature $r$ increases ($r_1 < r_2 < r_3$) due to geometrical constraint. (b) The Laplace pressure (negative) inside a capillary bridge is a monotonically increasing function of capillary bridge volume $V^{cb}$ (for acute contact angle much smaller than 90°). Here $D$ is set to zero, for nonzero small $D$ the trend is the same, see Figure S2 in the Appendix. (c) Analogy between Laplace pressure stabilizing meniscus size through inter-liquid diffusion (bottom) and hydrostatic pressure equilibrating liquid level through flow in communicating vessels (top).

Consider two spherical particles of radius $R$ with a small separation $D$ ($D \ll R$, Figure 2a) immersed in a two-phase liquid. Assume the particles are more lyophilic to liquid phase A, and phase A forms a pendular meniscus within the narrow slit. In 2D case, the Laplace pressure across the liquid interface of capillary bridge is simply $\Delta p = -\gamma_{AB}/r$, where $\gamma_{AB}$ is the interfacial tension between liquid phases A and B, and $r$ is the radius of curvature of the meniscus, which is dictated by the contact angle $\theta$ and the two-sphere surface geometry. The geometric confinement of the narrow slit between the two spheres prescribes that $r$ increases as the capillary bridge grows bigger in order to maintain the correct contact angle (Figure 2a). This means that the pressure inside the pendular meniscus monotonically increases as the capillary bridge grows bigger. To be general, we analyze 3D case (see Appendix for 2D analysis that exhibits qualitatively similar behavior). Using the commonly adopted circular approximation,[1, 2] i.e., assuming the meniscus profile along the axis of symmetry is described by a circular arc, the Laplace pressure inside the meniscus (internal pressure of liquid phase A with respect to external bulk pressure of liquid phase B) is given by Young-Laplace equation

$$p = \gamma_{AB} \left( \frac{1}{l} - \frac{1}{r} \right), \tag{4}$$

where $l$ and $r$ are respectively the azimuthal and meridional radii of the meniscus. Since usually $r \ll l$, the Laplace pressure is dominated by the radius $r$, just like in 2D case. Figure 2b plots the quantitative dependence of Laplace pressure on capillary bridge volume at different contact angles (see the Appendix for calculation details), which is practically the same as the exact solutions obtained without circular approximation by Heady and Cahn.[2] This plot emphasizes that capillary bridges defy the conventional coarsening process as usually observed during phase separation, in which Laplace pressure inside a droplet decreases as it grows bigger, thus bigger droplets grow at the expense of smaller ones. In capillary bridges, however, inter-liquid diffusion shrinks bigger bridges (with higher internal pressure) and feeds the growth of smaller ones (with lower internal pressure), which naturally creates a negative feedback for the growth of capillary bridges and automatically equilibrates their sizes. Such a self-stabilization mechanism of pressure-volume relationship with positive slope in capillary bridges (Figure 2b) resembles the role of Pascal's law ($\Delta p = \rho g \Delta h$) in communicating vessels to equilibrate liquid levels (Figure 2c).

The Laplace pressure across a curved interface leads to chemical potential shift in the liquid mixture in accord with the Gibbs-Duhem relation.[31] For the capillary bridges of liquid phase A ($c_1 \approx 1$)[25] in our simulation, the chemical potential $\mu_1 = \delta F / \delta c_1$ reflects the internal pressure. Figure 3 plots the chemical potential distribution inside the capillary bridges in the highlighted box shown in Figure 1 at time step 50,000. It clearly shows that $\mu_1$ is higher and nonuniform inside the bigger and unsymmetric capillary bridge as compared to that in the other two capillary bridges. It is this nonuniform chemical potential distribution that drives the bigger bridges to shrink and become symmetric. As a result, all capillary bridges interact with each other through inter-liquid diffusion and ultimately achieve equilibrium of uniform pressure and chemical potential. Such an effect is of critical importance since it implies an intrinsic self-stabilization mechanism among adjacent capillary bridges through diffusion. Therefore, the simulated formation of self-stabilized capillary bridges shown in Figure 1 is not by chance but a spontaneous process. In some systems, the phase separation process can be reversed through convenient temperature control.[11, 32] Many binary liquid systems, such as alchol-oil[33] and water-lutidine,[34] have a critical phase separation temperature close to room temperature, making it possible for the capillary bridge-reinforced colloidal crystal to be used at room temperature. The simulation thus indicates a potential route to *in-situ* firming of colloidal crystals, whose adhesion strength is evaluated in the next section.

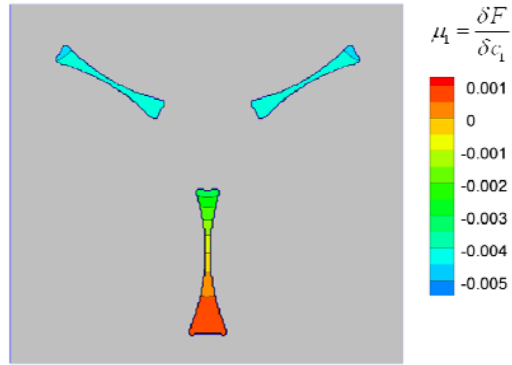

**Figure 3.** Contour plot of chemical potential $\mu_1 = \delta F / \delta c_1$ distributed inside the capillary bridges in the highlighted box shown in Figure 1 at time step 50,000, reflecting internal pressure of the capillary bridges.

3. Adhesion strength of capillary bridge-reinforced colloidal crystal

The total attractive capillary force between two spheres in contact is $F^{adh} = \omega \gamma_{AB} R$,[2] where $\omega$ is a coefficient; for capillary bridge volume $V^{cb} \sim 0.01 R^3$ or less, $\omega$ is approximately 5 for contact angle of 0° and 2 for contact angle 60°. Consider a closely packed 3D colloidal crystal of face-centered cubic symmetry, where all neighbouring spherical particles are bonded by capillary bridges. The average compressive stress contributed solely from capillary bridges is readily obtained as

$$\sigma_{ij} = -\sqrt{2} F^{adh} \delta_{ij} / 2R^2 ,  \quad (5)$$

where $\delta_{ij}$ is the Kronecker delta. Substituting $F^{adh} = \omega \gamma_{AB} R$ into Eq. (5) yields the tensile strength of the colloidal crystal as reinforced by the capillary bridges

$$\sigma^{TS} = \frac{\sqrt{2}\omega}{2} \frac{\gamma_{AB}}{R} . \quad (6)$$

The above formula indicates an inverse scaling of the tensile strength of the colloidal crystal with respect to particle radius, which is bounded by a theoretical limit of negative pressure that the liquid can sustain before cavitation (typically when the particle size reduces to nanometers).[9] Macroscopically, the adhesion caused by the self-stabilized uniform capillary bridges exhibits high degrees of isotropy and homogeneity. Assuming $\gamma_{AB} = 10^{-3} J/m^2$ and $R = 100\,nm$, then the tensile strength can reach the order of magnitude of $10^4\,Pa$, rendering sufficient strength to the colloidal crystals for many practical applications, such as 3D photonic crystals. Moreover, the strong capillary bridges would pull particles into more intimate contact and cause strengthening of van der Waals forces between the particles (see recent findings on the frozen colloid network after jamming at fluid interfaces[35, 36]). With appropriate particle surface properties, sintering[37] between particles can be controlled to further reinforce the structures if needed. A unique merit of the pendular capillary bridge reinforcement, in addition to its large strength, is that ideally the strong adhesion only increases equal normal contact stress between particles without introducing unwanted shear stress or torque to distort the original structure (as reported in experiment[11], the randomly packed microspheres did not undergo perceivable movement until coalescence of capillary bridges). Thus, the symmetry of the capillary bridge-reinforced colloidal crystal can be well preserved.

4. Volume fraction limit of capillary bridge liquid phase

The plot of internal pressure as a function of capillary bridge volume in Figure 2b is obtained based on the assumption of isolated capillary bridge, which breaks down if neighboring menisci meet together. For spherical particles closely packed in face-centered cubic crystal, coalescence of neighboring menisci occurs when the filling angle $\beta > 30°$ (see Figure S3 in the Appendix for critical capillary bridge volume). By calculating the critical capillary bridge volume $V^{cb\,*}$, the critical volume fraction of liquid phase A ($\Phi_A^*$) in the liquid mixtures is given by

$$\Phi_A^* = \frac{N_c f_p V^{cb\,*}}{2(1-f_p)V_p}, \tag{7}$$

where $V_p$ is the volume of a particle, $f_p$ is particle volume fraction, and $N_c$ is the coordination number of the crystal (each particle possess $N_c/2$ capillary bridges). It is worth noting that, as a capillary effect, the Laplace pressure causes a slight composition shift from the equilibrium composition of bulk phase diagram,[18, 19] which needs to be taken into account when using level rule to precisely control the volume of minor phase based on the binary phase diagram. Moreover, if $\Phi_A$ is below but very close to $\Phi_A^*$, there could still be a chance for capillary bridges to coalesce locally, especially in randomly packed colloidal structures due to local oversupplies of minor phase species and limited diffusion range, as demonstrated in Figure 4a. Once coalescence occurs, the liquid phase B is locally trapped and isolated from the particle surface, resulting in loss of the contact angle and interfacial tension. The droplets of phase B eventually shrink to disappear due to regular coarsening, leading to liquid clusters of phase A. Nevertheless, the internal pressure of the liquid clusters is determined by the curvature of the liquid interface of the clusters, thus equilibrium is reached between clusters and capillary bridges preventing overgrowth of clusters. However, such stabilized clusters are local defects since they locally weaken the adhesion strength of the capillary bridge-reinforced colloidal structures. In order to avoid local clustering and form uniform capillary bridges to reinforce imperfect colloidal structures, $\Phi_A$ should be chosen sufficiently below $\Phi_A^*$, as demonstrated in Figure 4b.

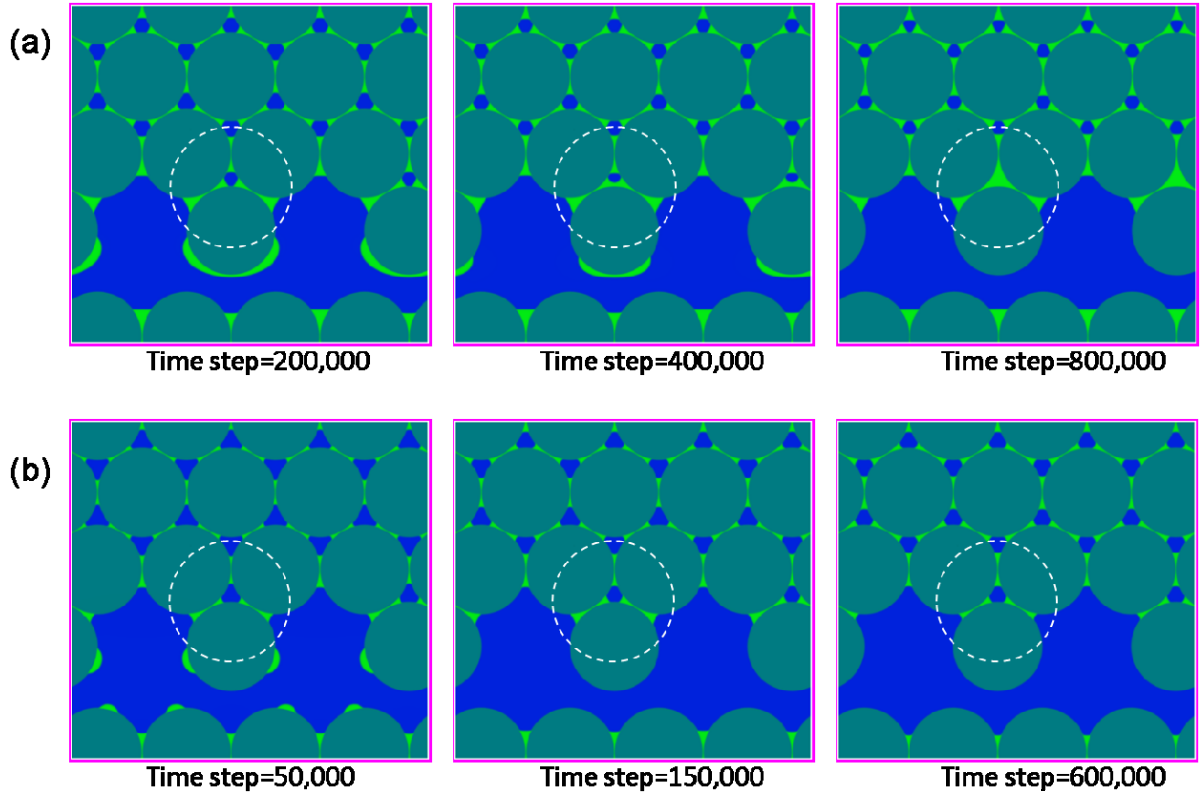

**Figure 4.** Simulated capillary bridge formation during liquid phase separation in imperfect colloidal crystal, where two particles are removed in addition to increased distance between rows 1 and 2 (colloid volume fraction $f_p$ thus reduces to 71%). $\Phi_A^*$ in this 2D configuration is about 0.20 (Note that $\Phi_A^*$ corresponding to configuration in Figure 1 is about 0.50). (a) $\Phi_A = 0.18$ close to $\Phi_A^*$, and capillary bridges inside the highlighted region coalesce and form clustered liquid phase A. (b) $\Phi_A = 0.10$ well below $\Phi_A^*$, thus coalescence of capillary bridges and liquid clustering are avoided, leading to uniform-sized capillary bridges.

5. Thermodynamic stability of capillary bridges

As shown in Figure 4a, a liquid cluster can also self-stabilize with respect to other capillary bridges as well as other liquid clusters. An interesting question is whether the uniform pendular state (as shown in Figure 1) is the lowest free energy state as compared to clustered states. In order to answer this question, we perform an imaginary relocation operation in which one block of liquid phase B between capillary bridges (see the triangular blue region in the last panel of Figure 1) is replaced with liquid phase A by equally extracting liquid A from all $N$ capillary bridges, and allow the system to relax and reach new equilibrium. Then the total system energies of the two states are compared, i.e., uniform capillary bridge state and nonuniform state with one liquid cluster. Similar operations are also performed for nonuniform states of two and three liquid clusters. Figure 5 shows these different states and compares their total system energies relative to the uniform state. The uniform capillary bridge state has the lowest energy, thus is thermodynamically stable. This conclusion holds when $\Phi_A$ is much smaller than $\Phi_A^*$.

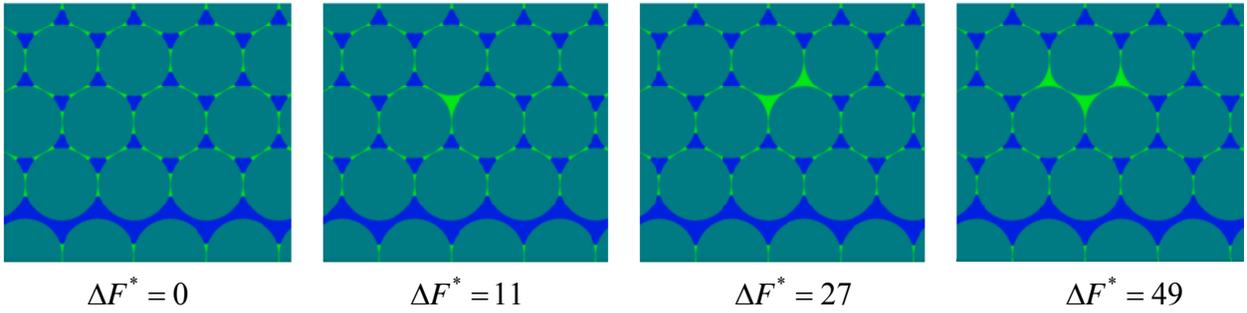

$\Delta F^* = 0$  $\quad\quad\quad\quad\quad$  $\Delta F^* = 11$  $\quad\quad\quad\quad\quad$  $\Delta F^* = 27$  $\quad\quad\quad\quad\quad$  $\Delta F^* = 49$

**Figure 5.** Comparison of simulated free energies of different liquid phase distributions in stable and metastable states. The same volume fraction $\Phi_A = 0.14$ of liquid phase A and particle configuration are used in all cases. The uniform capillary bridge state (first panel, without clustering) has the lowest energy.

The same conclusion of thermodynamic stability for uniform capillary bridge state can also be reached analytically. In above imaginary relocation operation, the replacement of the liquid phase B (of volume $\Delta V$) by liquid phase A reduces interfacial energy in that region by $-\Delta E^{int}$, whereas there is interfacial energy increase $\delta E^{cb}$ associated with each capillary bridge due to its volume change $\delta V^{cb} = -\Delta V/N$, where $N$ is the number of capillary bridges, while the bulk energy change is zero to the first-order approximation due to diffusive equilibrium in the original state. The excess free energy of the whole system during this imaginary process is thus mainly from interfacial energy changes

$$\Delta F = -\Delta E^{int} + N \cdot \delta E^{cb}, \tag{8}$$

where

$$N \cdot \delta E^{cb} = N \cdot \frac{dE^{cb}}{dV^{cb}} \delta V^{cb} = -\frac{dE^{cb}}{dV^{cb}} \Delta V. \tag{9}$$

The first term in Eq. (8) is negative, while the sign of the second term depends on $dE^{cb}/dV^{cb}$. The dependence of capillary bridge interfacial energy $E^{cb}$ on the bridge volume $V^{cb}$ is plotted in Figure 6, which shows that $dE^{cb}/dV^{cb}$ is mostly negative (except for large contact angle) with absolute value rapidly increased for small $V^{cb}$. This implies that the second term in Eq. (8) is positive and dominates over the first term in $\Delta F$ for small $V^{cb}$. Thus, when capillary bridge volume is small, $\Delta F > 0$, and the uniform capillary bridge state has the lowest energy than nonuniform states with clustering of liquid phase A. This conclusion is in agreement with the simulation results shown in Figure 5.

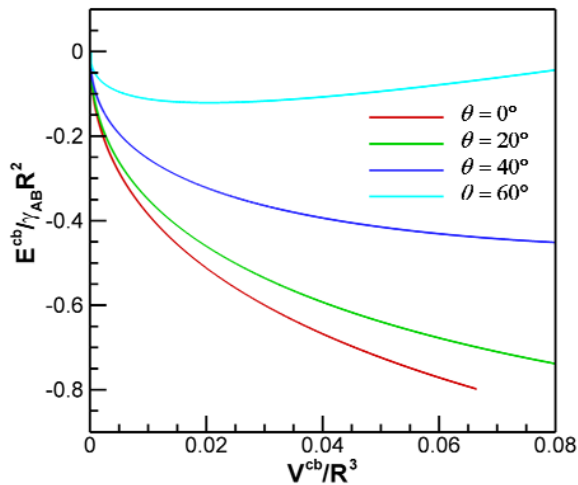

**Figure 6.** Dependence of interfacial energy associated with a capillary bridge on the capillary bridge volume. (see Appendix calculation details)

6. Simulation of capillary bridge formation in other engineered colloid assemblies

The self-stabilization mechanism (communicating vessels effect) makes the capillary bridges very tolerant of colloidal structure disorders as long as $\Phi_A$ is well below $\Phi_A^*$ to avoid coalescence of neighboring capillary bridges. Such a mechanism is also able to stabilize capillary bridges in other compact colloid assemblies. The essential condition for the self-stabilization mechanism is that the internal pressure is a monotonically increasing function of the bridge volume, hence the particles are not restricted to be identical spheres. Even if there exist different particle geometries,[1] as long as the Laplace pressure falls into a common range, equilibrium can also be achieved among the capillary bridges. Such behaviors are further demonstrated in Figures 7 and 8. Figure 7 shows the simulated capillary bridges formed between neighboring colloidal particles in a Saturn-ring-like superstructure.[38] The center sphere has different radius, thus the final stable capillary bridges assume two different sizes. Just as in colloidal crystals, the capillary bridges in this microsphere assembly solely increase normal contact stress between the particles, thus the microstructure and symmetry of the assembly are well maintained. Figure 8 shows the simulated capillary bridges formed in an interstitial colloidal crystal consisting of spheres of two different sizes, where stable uniform capillary bridges are also obtained. In all these cases, the capillary bridges reinforce the otherwise fragile colloidal structures.

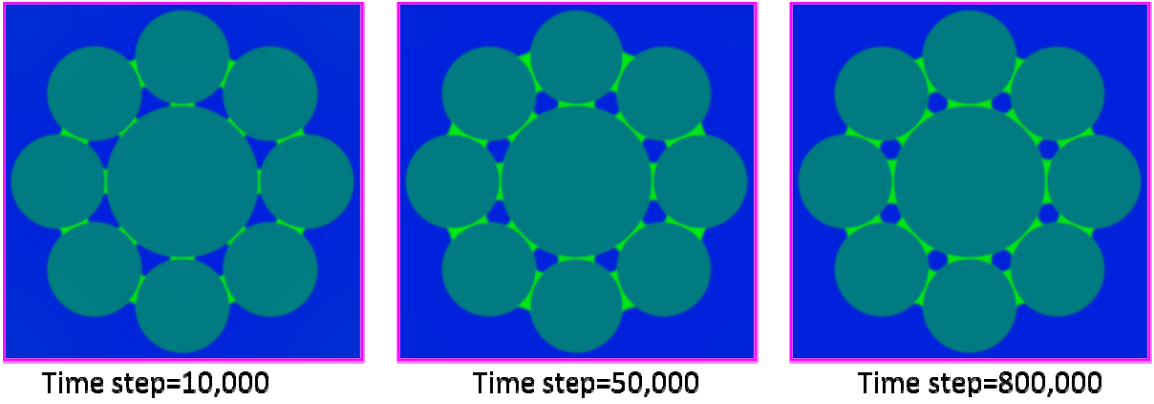

**Figure 7.** Simulated capillary bridge formation and self-stabilization in saturn-ring-like colloidal superstructure. The radii of the center and satellite spheres are $112l$ and $70l$, respectively. $528\times528$ mesh is used. $\Phi_A = 0.07$.

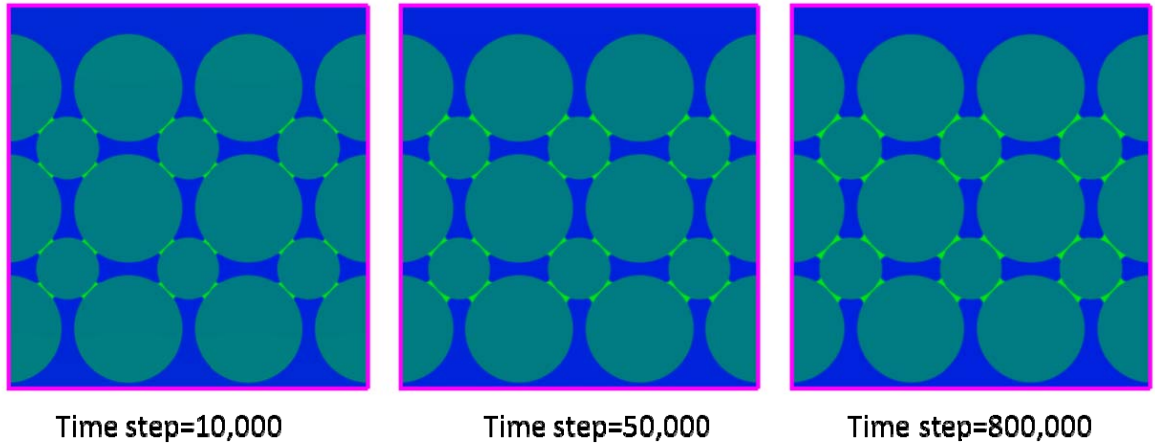

**Figure 8.** Simulated capillary bridge formation and self-stabilization in an interstitial colloidal crystal. The radii of different-sized spheres are $112l$ and $70l$, respectively. $\Phi_A = 0.07$. $768\times816$ mesh is used.

**CONCLUDING REMARKS**

Computer simulations of capillary bridge formation in various closely packed colloidal assemblies through liquid phase separation reveal a self-stabilization mechanism that operates through diffusive equilibrium of two-phase liquid morphologies. Such mechanism renders desired microstructural stability and uniformity to the capillary bridges, which can be used to reinforce the colloidal microstructures with sufficient adhesion strength. This study sheds light on a promising route to *in-situ* (in-liquid) firming of fragile colloidal crystals and other compact colloidal microstructures via spontaneous formation of uniform and stable capillary bridges through liquid phase separation, where the capillary bridges bond colloidal particles as mortar bonds bricks.

**ACKNOWLEDGEMENTS**

We thank Professor Yongmei M. Jin for her help with the development of DIFA code. Support from NSF under Grant No. DMR-0968792 is acknowledged. Simulations were performed on TeraGrid supercomputers.

**APPENDIX: Calculation details of capillary bridges**

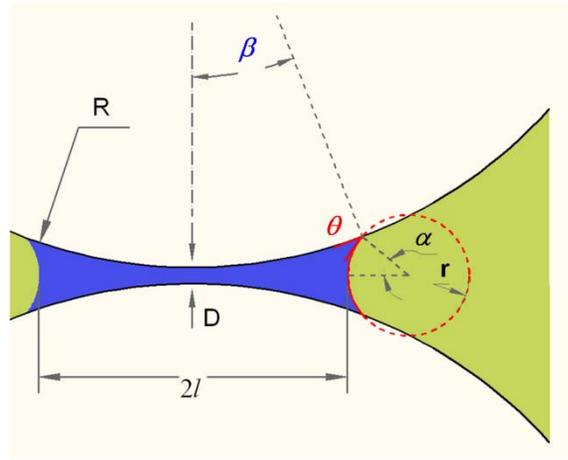

**Figure S1.** Definition of geometrical parameters for a meniscus in a slit between two spheres.

By adopting the circular approximation, the influence of geometrical confinement is straightforward: in order to maintain correct contact angle, the radius of meridional curvature is restricted to be (see Figure S1 for the definition of geometrical parameters)

$$r = \frac{2R(1-\cos\beta)+D}{2\cos(\theta+\beta)}, \quad (S1)$$

and the azimuthal radius is

$$l = R\sin\beta - r(1-\cos\alpha), \quad (S2)$$

with $\alpha = \frac{\pi}{2}-(\beta+\theta)$ (shown in Figure S1) defined for convenience.

The pressure inside the meniscus can be thus calculated according to the Young-Laplace equation

$$p = \gamma_{AB}\left(\frac{1}{l}-\frac{1}{r}\right). \quad (S3)$$

The volume of a capillary bridge can be calculated as

$$V^{cb} = V^{rot} - 2V^{cap}, \tag{S4}$$

where

$$V^{rot} = 2\pi[r(l+r)^2 \sin\alpha - r^2(l+r)(\sin 2\alpha + 2\alpha)/2 + r^3(\sin\alpha - \sin^3\alpha/3)], \tag{S5}$$

and

$$V^{cap} = \frac{\pi}{3}R^3(2+\cos\beta)(1-\cos\beta)^2. \tag{S6}$$

The relation between pressure and the volume of capillary bridge is thus obtained from Eqs. (S3)-(S6). Instead of solving for an explicit function, it is easier to obtain the relationship curve as a parametric function through the parameter $\beta$ (we plot it by Matlab). In Fig. 1b, we set $D=0$, while the same trend of monotonic increase is found for nonzero $D$, as shown in Figure S2. Also, the volume of a capillary bridge corresponding to different filling angles can be plotted (as shown in Figure S3).

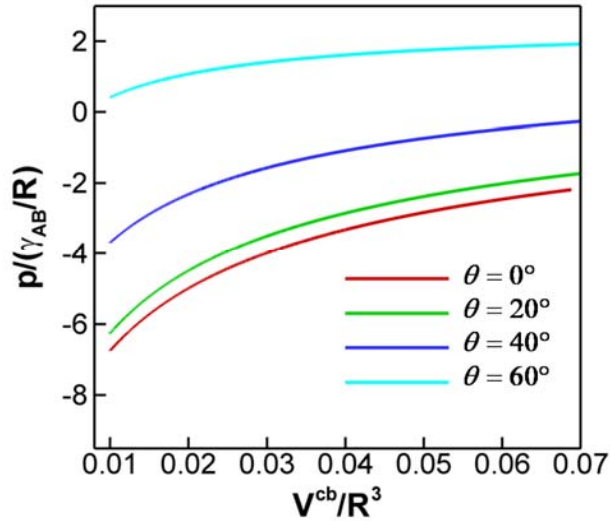

**Figure S2.** For acute contact angles, Laplace pressure still monotonically increases with increasing volume of a capillary bridge for small nonzero gap distance ($D = 0.02R$).

The interfacial areas between particle/phase A $S_{AP}$ and phase-A/phase-B $S_{AB}$ can also be obtained by

$$\begin{aligned} S_{AB} &= 4\pi r[l\alpha + r(\alpha - \sin\alpha)] \\ S_{AP} &= 4\pi R^2(1-\cos\beta) \end{aligned}. \tag{S7}$$

The interface energy associated with an individual capillary bridge is

$$\Delta E^{cb} = S_{AB}\gamma_{AB} + S_{AP}(\gamma_{AP} - \gamma_{BP}) = (S_{AB} - S_{AP}\cos\theta)\gamma_{AB}, \tag{S8}$$

so the dependence of $\Delta E^{cb}$ on the volume $V^{cb}$ can be calculated (Fig. 6 in the main text).

In two-dimensional cases, the calculation is much easier. Following the above approach, the pressure-volume relationship is obtained and shown in Figure S4, with the same trend as in 3D case.

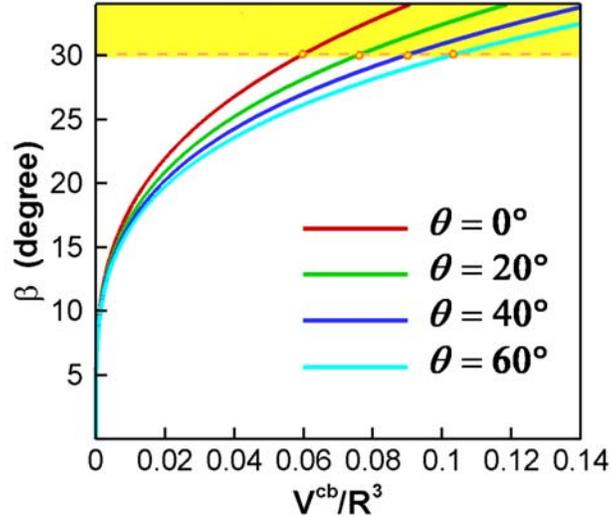

**Figure S3.** Dependence of filling angle $\beta$ on capillary bridge volume at different contact angles. Yellow zone indicates coalescence of neighboring menisci for fcc or hcp crystals.

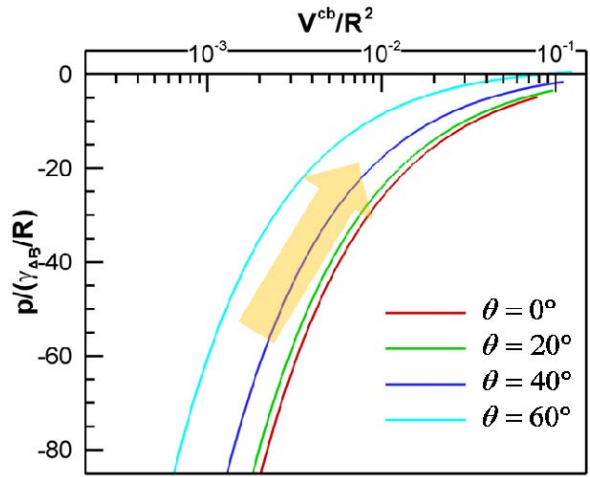

**Figure S4.** The Laplace pressure of a capillary bridge (2D) is also a monotonically increasing function of the capillary bridge volume, $V^{cb}$ (for cases of acute contact angle, well smaller than 90°).